\documentclass[twocolumn]{aastex63}

\usepackage{amsmath}	
\usepackage{lineno}
\usepackage{bm}
\usepackage[normalem]{ulem}
\usepackage{float}
\usepackage{soul}

\newcommand{\Mdot}{\dot{M}}
\newcommand{\Mdotacc}{\dot{M}_{\rm acc}}

\newcommand{\Mdotedd}{\dot{M}_{\rm Edd}}
\newcommand{\mesa}{\texttt{MESA}}
\newcommand{\yd}{y_{\rm d}}
\newcommand{\h}{^1{\rm H}}

\newcommand{\src}{SAX~J1808.4--3658}
\newcommand{\iso}[2]{^{#1}{\rm #2}}
\newcommand{\maestro}{\texttt{MAESTRO}}
\newcommand{\maestroex}{\texttt{MAESTROeX}}

\shorttitle{Hydrodynamics of proton-ingestion flashes}
\shortauthors{Guichandut et al.}

\begin{document}
\title{Hydrodynamical simulations of proton ingestion flashes in Type I X-ray Bursts}

\correspondingauthor{Simon Guichandut}
\email{simon.guichandut@mail.mcgill.ca}

\author{Simon Guichandut}
\affiliation{Department of Physics and Trottier Space Institute, McGill University, 3600 rue University, Montreal, QC, H3A 2T8, Canada}

\author{Michael Zingale}
\affiliation{Department of Physics and Astronomy, Stony Brook University, NY 11794-3800, USA}

\author{Andrew Cumming}
\affiliation{Department of Physics and Trottier Space Institute, McGill University, 3600 rue University, Montreal, QC, H3A 2T8, Canada}

\begin{abstract}
We perform the first multidimensional fluid simulations of thermonuclear helium ignition underneath a hydrogen-rich shell. This situation is relevant to Type I X-ray bursts on neutron stars that accrete from a hydrogen-rich companion. Using the low-Mach number fluid code \maestroex{}, we investigate the growth of the convection zone due to nuclear burning, and the evolution of the chemical abundances in the atmosphere of the star. We also examine the convective boundary mixing processes that cause the evolution to differ significantly from previous one-dimensional simulations that rely on mixing-length theory. We find that the convection zone grows outward as penetrating fluid elements cool the overlying radiative layer, rather than directly from the increasing entropy of the convection zone itself. Simultaneously, these flows efficiently mix composition, carrying carbon out of and protons into the convection zone even before contact with the hydrogen shell. We discuss the implications of these effects for future modeling of these events and observations.
\end{abstract}

\keywords{X-ray bursts --- Neutron stars --- Convection --- Hydrodynamics}

\section{Introduction}\label{sec:intro}
Type I X-ray bursts are the result of thermonuclear runaways on the surface of accreting neutron stars in low-mass X-ray binaries \citep{Lewin.Paradijs.ea1993,Galloway.Keek2021,Strohmayer.Bildsten2006}. As the most frequent transient in the high-energy sky with over 7000 bursts cataloged \citep{Galloway.Zand.ea2020}, these ``bursts'' can help us understand the properties of the neutron star surface. The challenge is to properly model the physics of nuclear burning, fluid motions, radiation, and their interaction.

A critical mechanism that controls the evolution of these bursts is convection. Even though ignition happens only a few tens of meters below the star's surface, the density of the envelope is so large that the fluid cannot cool by radiation only. The burning fuel will therefore generate a convection zone that will grow toward the surface in a time shorter than the thermal time. Many studies have described this convection zone and its evolution. \citet{Joss1977} derived the basic timescales for radiation and convection, and showed that convection could not reach all the way to the photosphere. Initial time-dependent calculations made the simplifying assumption that regions unstable to convection are perfectly adiabatic and instantaneously mix entropy \citep{Joss1978,Hanawa.Sugimoto1982}. Many simulations of bursts were performed using stellar evolution codes such as \texttt{KEPLER} \citep{Wallace.Woosley.ea1982,Woosley.Heger.ea2004}, \texttt{SHIVA} \citep{Jose.Moreno.ea2010}, and \mesa{} \citep{Paxton.Bildsten.ea2011,Yu.Weinberg2018}. These codes have in common that they treat convection with some variation of mixing-length theory \citep[MLT, e.g.][]{Cox.Giuli1968,Henyey.Vardya.ea1965}, in which it is assumed that convective parcels travel a fixed fraction of the local pressure scale height, at the velocity required to transport the heat flux.

Meanwhile, few studies have been dedicated to modeling the multidimensional hydrodynamic nature of convection within bursts. Numerically, this is a challenge of timescales. Burning and convection take place over many scale heights, and one therefore needs to account for the compressibility of the fluid. This introduces sound waves into the system, which travel on a sound crossing time of ${\sim}10\,\mu$s, much shorter than the tens of seconds burst duration, and the computational cost of the simulations easily becomes prohibitive. But since bursts proceed as subsonic deflagrations \citep{Wallace.Woosley.ea1982}, an alternative approach is to model the fluid in the low-Mach number approximation \citep[see][and references therein for the origins of this method]{Almgren.Bell.ea2006}. This was first applied to X-ray bursts by \citet{Lin.Bayliss.ea2006} who performed two-dimensional (2D) simulations of a pure helium-burning layer and growing convection zone, and confirmed that the Mach number remained small ($\lesssim 15$\%) throughout, justifying the use of the low-Mach method. \citet{Malone.Nonaka.ea2011} used the dedicated low-Mach number hydrodynamics code \maestro{} \citep{Nonaka.Almgren.ea2010} to model a similar type of burst. \citet{Malone.Zingale.ea2014} then modeled a burst from a uniform mixture of hydrogen and helium, a more common scenario for accreting neutron stars in bursting sources. Both works found that a very high spatial resolution (few centimeters per zone or even less, depending on the temperature sensitivity of the nuclear reactions) was needed to obtain numerical convergence. 

Another possibility for the composition of the envelope prior to the outburst is a layered structure, where a hydrogen-rich shell sits on top of a pure helium layer. In sources where the neutron star is accreting solar-like material (primarily H with mass fraction $X_0\sim0.7$, then He, and some CNO elements with $Z_{\rm CNO}$ of a few percent), at high enough accretion rates $\Mdotacc$ such that the temperature $T$ exceeds $8\times 10^7$ K, hydrogen will continuously burn via the hot CNO cycle, and run out at the depletion depth
\begin{equation}\label{eq:yd}
    \yd = 2.7\times 10^7\,\rm{g}\;\rm{cm}^{-2}\left(\frac{\Mdot_{\rm acc}}{0.01 \Mdotedd}\right)\left(\frac{0.02}{Z_{\rm CNO}}\right)\left(\frac{X_0}{0.7}\right)\,,
\end{equation}
as shown by \citet{Cumming.Bildsten2000}. Here $\Mdot_{\rm Edd}$ is the Eddington accretion rate for a neutron star with a 12 km radius accreting gas with $X_0=0.7$. The column depth $y(r)\equiv\int_r^\infty \rho(r')dr'$ measures the amount of mass above a radial coordinate $r$. 

Recently, in \citet{Guichandut.Cumming2023}, we suggested that the lightcurves of bursts which ignite at $y>y_d$ could be used to test the radial extent of the convection zone. In particular, we showed that the peculiar observation of a ${\approx}\,0.7\,$s pause in the rise of a burst from \src{} \citep{Bult.Jaisawal.ea2019} was consistent with the ejection of a hydrogen shell. For the pause to be so short, this shell would have had to be eroded by convection. We ran one-dimensional calculations using the stellar evolution code \mesa{} \citep{Paxton.Bildsten.ea2011} and verified that convection could indeed extend to the required column depths of $y_{\rm conv}{\sim}10^4{-}10^5$ g cm$^{-2}\ll \yd$. However, our results were sensitive to assumptions made about convection in the code. One reason for this is that MLT is not appropriate for situations where the thermodynamics of the gas are changing on timescales similar to the convective turnover time. In these bursts, a convection zone in the helium layer will grow and eventually ``collide'' with the hydrogen-rich shell above, which will introduce new fuel in the form of free protons which quickly burn and rapidly change the nature of the convection. This is an example of a ``proton-ingestion flash'', which also occur for example in evolved metal poor stars \citep[see e.g.][and references therein]{Herwig.Pignatari.ea2011}. The limitations of MLT have been noted for these events as well, and recent works have instead focused on fluid simulations \citep{Herwig.Woodward.ea2014}.

To understand how convection proceeds in X-ray bursts with proton-ingestion events, we must model the multidimensional fluid. In this paper, we use the low-Mach number hydrodynamics code \maestroex{} \citep[the successor to \maestro; ][]{Fan.Nonaka.ea2019} to evolve the same burst which we previously studied in \citet{Guichandut.Cumming2023}. As in \citet{Malone.Nonaka.ea2011,Malone.Zingale.ea2014}, we consider a 2D box in the plane-parallel neutron star atmosphere on the verge of thermonuclear ignition. We briefly describe the numerical method and how our initial model is created in Section \ref{sec:methodology}, and present the results of the simulations in Section \ref{sec:results}. We conclude in Section \ref{sec:discussion}.

\section{Methodology}\label{sec:methodology}
\subsection{Numerical method}
The basic idea of the low-Mach number method is to decompose the pressure into base-state plus perturbative components,
\begin{equation}
    p=p(\bm{x})=p_0(r)+\pi(\bm{x})\,,
\end{equation}
where $\bm{x}$ is the cartesian position vector and $r$ is the radial coordinate in the star. The base state is assumed to be in hydrostatic equilibrium, i.e.
\begin{equation}
    \nabla p_0=-\rho_0 g\bm{\hat{r}}\,,\label{eq:HSE}
\end{equation}
where $\rho_0(r)$ is the base state density. It can be shown \citep{Almgren.Bell.ea2006} that if the Mach number $M=\vert \bm{U}\vert/c_s$ ($c_s$ being the local speed of sound) is small, then the pressure perturbations $\pi/p_0$ are of order $M^2$. Thus, we can ignore $\pi$ in linking the pressure to the density via the equation of state, which effectively decouples the two and filters out sound waves. For full details on the equations being solved and the algorithmic implementation, see \citet{Fan.Nonaka.ea2019}.

As in previous papers using \maestro{} to model bursts, we use the Helmholtz equation of state \citep{Timmes.Swesty2000}, which has pressure contributions from ideal gas, radiation, and degenerate electrons. Our reaction network is equivalent to \mesa{}'s ``CNO\_extras'' network \citep{Paxton.Bildsten.ea2011}, with inert $\iso{56}{Fe}$ added.  The complete set of reactions linking 20 isotopes from $\iso{1}{H}$ to $\iso{24}{Mg}$ is assembled using \texttt{pynucastro} \citep{Smith-Clark.Johnson.ea2023}.

\subsection{Initial model}\label{sec:initial_model}
\begin{figure}
    \centering
    \includegraphics{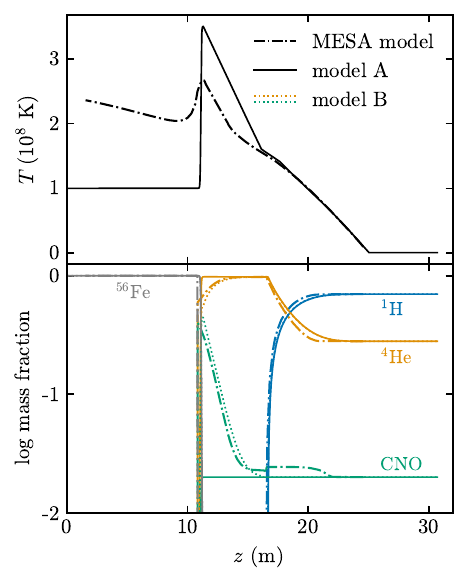}
    \caption{Initial model from \mesa{} (dash-dotted lines), and the initial model A fed into the fluid simulations (solid lines). The top panel shows the temperature profile as a function of height. The bottom panel shows the composition profiles of the main chemical species, where the green color combines all CNO species. The offset in the position of the solar layer is due to the temperature kick given to the initial fluid model. An alternate initial model (model B) including deep carbon is shown in the dotted lines for CNO and helium mass fractions. See Appendix~\ref{sec:app_initial_model} for details.}
    \label{fig:initial_models}
\end{figure}

As in \citet{Malone.Zingale.ea2014}, we create an initial model based on a 1D stellar evolution calculation, in this case our previous \mesa{} study of this type of burst in \citet{Guichandut.Cumming2023}. There, we considered a neutron star with a mass of $1.4 M_\odot$, radius $12$ km, and surface gravity of $1.29\times10^{14}$ cm s$^{-2}$ (ignoring relativistic corrections). We first accreted a large column of inert iron to form the bottom of the atmosphere, then let it accrete a solar mixture of fuel ($\iso{1}{H}$, $\iso{4}{He}$ and $\iso{12}{C}$ with mass fractions $X=0.7$, $Y=0.28$ and $Z=0.02$) until ignition. This \mesa{} model is plotted with dash-dotted lines in Figure~\ref{fig:initial_models}, in terms of its temperature and composition.

In practice, it is difficult to directly port a \mesa{} model to a fluid simulation code, because the former's grid is by mass coordinate while the latter is spatial, and due to small differences in the EOS of both codes. This is why we instead opt to construct a toy model which is defined by a small set of physical parameters. 

The full description of our model is presented in Appendix \ref{sec:app_initial_model}. Briefly, our atmosphere is composed of solar-like fuel at the top, then pure helium and CNO past the hydrogen depletion depth, then pure iron. At the bottom of the fuel layer (just above the iron) is an isentropic zone which becomes convectively unstable once the simulation begins. The temperature profile above is in radiative equilibrium, and depends on the flux from nuclear burning. We set the temperature of the iron layer to an artificially small value so as to create a ``buoyancy wall'' that suppresses downward penetrating flows and inhibits mixing of the iron into the fuel above. Lastly, we ``kick'' the model by increasing the temperature of the isentropic zone in order to accelerate the evolution toward the thermonuclear runaway. This ``model A" is shown with solid lines in Figure~\ref{fig:initial_models}.

This model is simpler than the \mesa{} model in two main aspects. First, we assume that the initial CNO species are $\iso{14}{O}$ and $\iso{15}{O}$ only. This is valid for $y<\yd$ where hot CNO burning is taking place and the rate-limiting step is the oxygen $\beta$-decays. (Indeed, our \mesa{} simulations confirmed that most of the accreting CNO converts to oxygen before hitting the depletion depth). However, for $y<\yd$, there are no more protons to sustain the reactions and the oxygens should decay to $\iso{15}{N}$, $\iso{14}{N}$, then $\iso{13}{C}$ products \citep[see Figure 2 in ][]{Guichandut.Cumming2023}. We do not expect this simplification to affect the results because these species only represent a few percent of the mass of the fuel, and because burning is initially dominated by triple-$\alpha$, and then by its products, as we will see later.

Second, we start in a state where the helium is fully unburned, whereas in reality, over the course of days of accretion and stable burning, some of the helium would have already converted to $\iso{12}{C}$. This carbon becomes the dominant CNO species in the helium layer, reaching a peak mass fraction of ${\approx}~0.4$ at the bottom (see Figure \ref{fig:initial_models}). To investigate the impact that this carbon has on the burst evolution, we create an alternate initial model, ``model B", with the same temperature profile but with a similar total mass of carbon as the \mesa{} model. The adjusted carbon and helium mass fractions for this other model are shown with dotted lines in the bottom panel of Figure~\ref{fig:initial_models}.

As in \citet{Malone.Zingale.ea2014}, we configure a ``sponge'' region at the top of the model. Below a density $\rho=\rho_{\rm cutoff}=10^3$ g cm$^{-3}$, we hold the density constant. Within this region, \maestroex{} solves the anelastic velocity constraint, which effectively suppresses large spurious velocities which would otherwise make the calculation diverge. The sponge initially starts at $z\approx 22$ m, but moves upward, following the density, as the atmosphere expands during the burst.

\section{Results}\label{sec:results}

We ran simulations, starting from the initial model presented in the last section, at resolutions of 12, 6, 3 and 1.5 cm per zone. The grid is square with a width of 3072 cm. This being much less than the radius of the star, the atmosphere is effectively plane-parallel, with a constant gravity $g=1.29\times 10^{14}$ cm$^{2}$ s$^{-1}$. The boundary conditions are periodic on the sides, outflow at the top, and slip wall (vanishing vertical velocity and tangential velocity gradient) at the bottom of the box. The density and enthalpy are held constant at the bottom and at the top. We do not include thermal diffusion as it is much slower than heat transport via convection at the depths we are considering\footnote{\citet{Malone.Nonaka.ea2011} showed that thermal diffusion only had minor effects on the evolution of convection, which is what we are focusing on. We also verified that the thermal diffusion timescales were long (${\sim}0.1$~s or more) for the largest temperature gradients found in our simulations.}.

In Section \ref{subsec:convergence}, we compare the general evolution of the different simulations and discuss convergence. In Section \ref{subsec:evolution} we consider a single simulation to analyze the evolution of convection and burning throughout the burst. In Section \ref{subsec:boundary}, we delve into the details of the mixing near the convective boundary. These three sections all consider model A. In Section \ref{subsec:carbon}, we look at a simulation of a more explosive version of this burst using model B.

\subsection{Resolution dependence and convergence}\label{subsec:convergence}

\begin{figure}[ht]
    \centering
    \includegraphics[width=\columnwidth]{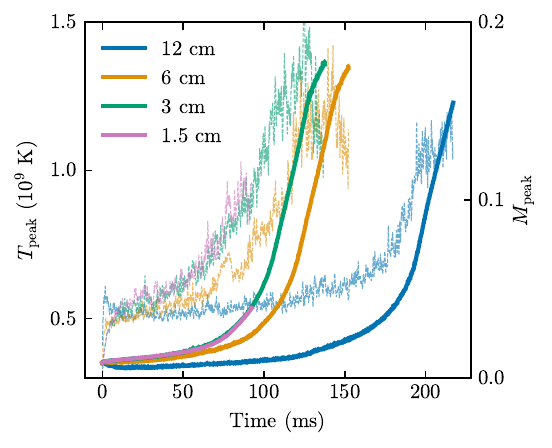}
    \caption{Peak values of the temperature (left axis, solid lines) and mach number (right axis, dashed lines) as a function of time for the same burst at different resolutions.}
    \label{fig:peaks}
\end{figure}

In order to assess convergence, we plot the instantaneous maximum values in the grid of the temperature and Mach number in Figure~\ref{fig:peaks}. Both peaks are tracers of the evolution of the burst in time. The temperature curve clearly shows a transition from a slow increase to a full runaway. As we will see later, this transition is related to the collision of the convection zone with the overlying hydrogen-rich shell.

The evolution of the burst in time clearly depends on the resolution. At greater resolutions (smaller grid spacings), the burst evolves faster. A few factors contribute to this, which were previously observed and discussed in \citet{Malone.Nonaka.ea2011} and \citet{Malone.Zingale.ea2014}. First, since the nuclear burning is very sensitive to temperature, under-resolving the peak of the temperature profile will underestimate the energy generation. Second, lower resolution results in spurious large velocities which overshoot the convection zone, taking heat away from it and inhibiting its growth. In our case, this second factor is particularly important as convection is needed to bring additional fuel into the mixture. 

The simulations appear to converge at a grid spacing of 3 cm, as decreasing it to 1.5 cm does not significantly change the evolution. For comparison, \citet{Malone.Nonaka.ea2011} obtained convergence at a resolution of 0.5 cm; however, the X-ray burst under study had different initial conditions (notably a much larger initial temperature). Despite convergence demonstrating the robustness of the \maestroex{} calculations, our simulations do reach appreciable fractions of the mach number ($\gtrsim$15\%). Shortly thereafter, the calculation becomes unstable, and the simulations stop. At this point, the thermonuclear runaway is still underway and the flux has not had time to escape the atmosphere. Therefore, the full evolution of Type I X-ray bursts in the low-Mach approximation remains intractable. We discuss this problem further in Section \ref{sec:discussion}.

\subsection{Evolution of convection}\label{subsec:evolution}

\begin{figure*}[h]
    \centering
    \includegraphics[width=\textwidth]{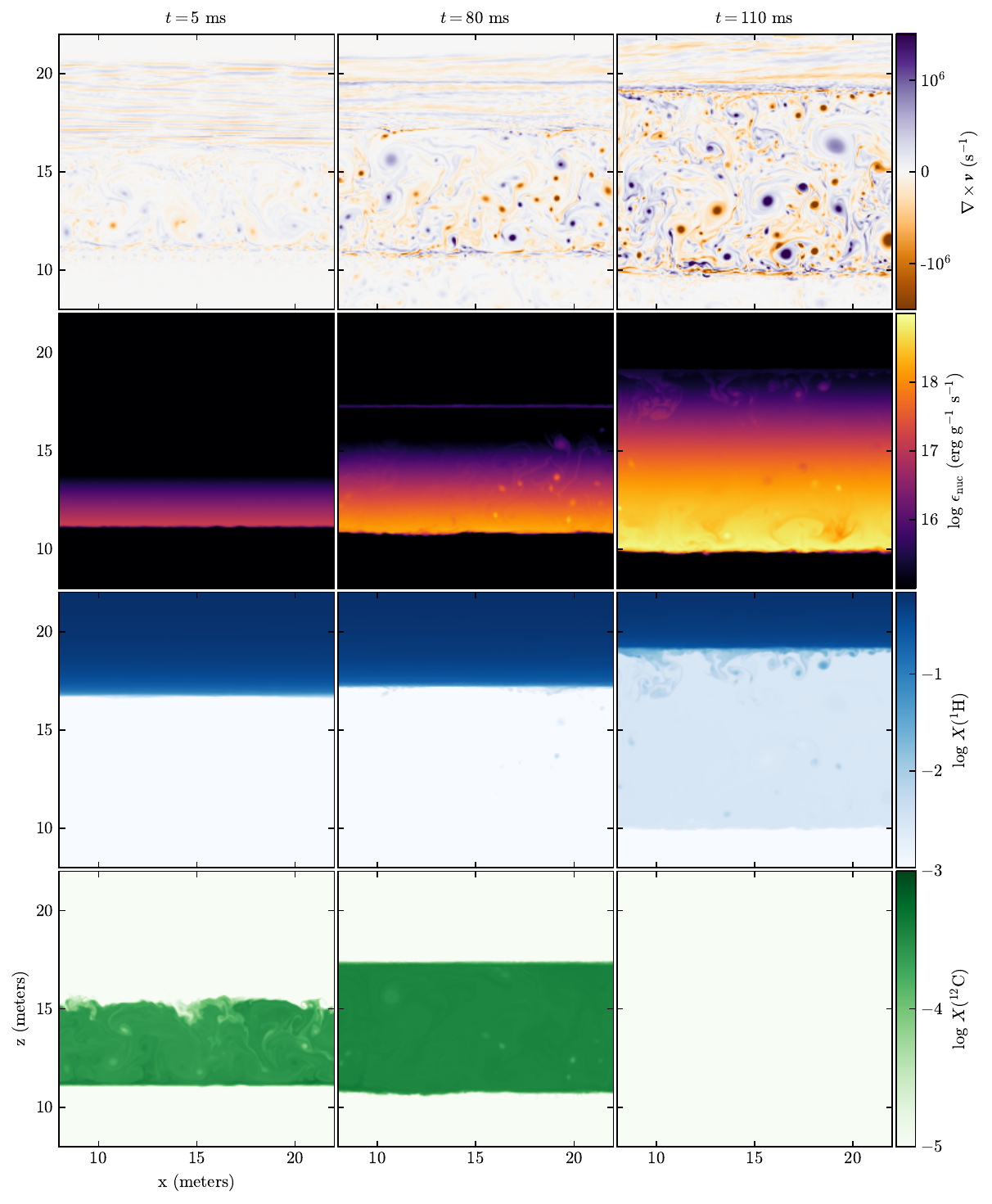}
    \caption{Flow, burning, and composition at different times in the 3 cm \maestroex{} simulation of model A. From top to bottom: vorticity, burning rate, mass fraction of hydrogen, mass fraction of carbon. Note the different scales in the bottom two rows. The left column is during the initial growth of the convection zone in the helium layer, the middle column is roughly at the collision with the hydrogen layer, and the right column is later during the expansion into the hydrogen. An animated version of this figure is available in the online journal.}
    \label{fig:panels}
\end{figure*}

\begin{figure*}
    \centering
    \includegraphics[width=\textwidth]{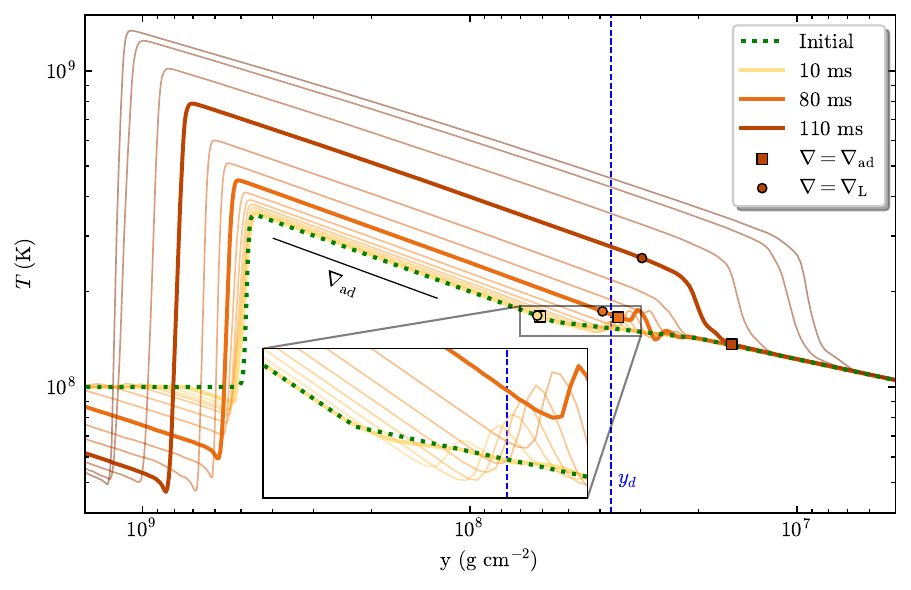}
    \caption{Evolution of the temperature as a function of column depth (note that the x-axis is reversed), at intervals of 10 ms. The colors get darker as time increases. The labeled and thicker lines are the three times from Figure~\ref{fig:panels}. The circle and square markers show the convective boundaries according to the Schwarzschild and Ledoux criteria respectively. The vertical dotted line is the depletion depth, i.e. the depth of the hydrogen layer. As convection approaches this line, the Schwarzschild and Ledoux boundaries begin to separate. The inset zooms into the evolution near the hydrogen boundary.}
    \label{fig:T_evolution}
\end{figure*}

We now look in depth at the evolution of the 3 cm model. We identify three ``special'' times for the three stages which we explore in this section. In Figure~\ref{fig:panels}, we plot snapshots of the fluid at these times in terms of its vorticity, nuclear burning, hydrogen and carbon mass fractions.

\textit{Growth in the He layer.} We first start the simulation by applying a gaussian random noise perturbation of 1 part in 1000 to every temperature in the grid. This triggers initial flows, which rapidly merge into a convective region in the initial isentropic zone. In the top-left panel of Figure~\ref{fig:panels}, convection can be seen by the presence of large vortices, a distinct feature of 2D convection. These vortices travel up and down the convective region, entrained by the greater overturning convective motions. With convective velocities of $\sim$km s$^{-1}$, the convective turnover times at this point are on the order of milliseconds. Burning is initially confined to the bottom of the convection zone. Triple-$\alpha$ reactions convert helium into carbon and release heat into the convective region, causing it to expand. After ${\sim}10$~ms, carbon becomes fully mixed in the convective region, uniformly increasing in mass fraction as helium burns.

In Figure~\ref{fig:T_evolution}, we show the evolution of the temperature profile\footnote{Note that here and in upcoming figures, the profiles of a quantity with depth are obtained via horizontal averaging.}. Throughout this initial evolution, the temperature gradient of the convection zone remains adiabatic ($\nabla\equiv d\ln T/d\ln p$ is close to $\nabla_{\rm ad}$, as indicated in the figure). As a result, the top of the convection zone is roughly located at the intersection between this gradient and the initial radiative profile (but not exactly, as we will discuss in Section \ref{subsec:boundary}). Because the convection zone and overlying stable He zone have similar mean molecular weights, the Schwarzschild and Ledoux\footnote{We evaluate the Ledoux gradient $\nabla_{\rm L}$ as in \mesa{}, see Section 3.3 of \citet{Paxton.Cantiello.ea2013}.} criteria agree on the location of the convective boundary, which can be seen in Figure~\ref{fig:T_evolution} with the square and circle markers lying on top of each other.

\begin{figure}[ht]
    \centering
    \includegraphics[width=\columnwidth]{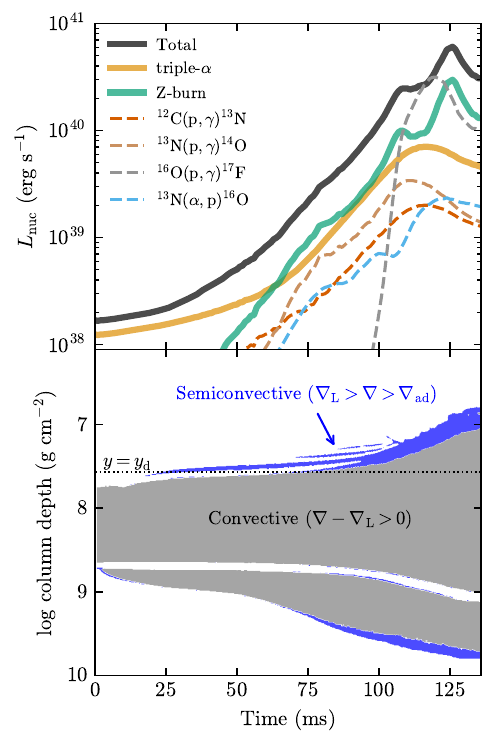}
    \caption{\textit{Top}: evolution of the nuclear luminosity (Equation~\ref{eq:Lnuc}) with time, for helium-burning and metal burning reactions (including CNO), and the total. The production of the most important CNO rates is plotted in dashed lines. \textit{Bottom}: size of the convective and semiconvective regions as a function of time. The dotted line shows the depletion depth.}
    \label{fig:Lnuc}
\end{figure}

\textit{Collision with the H shell.} After about 50 ms, overshooting motions above the convective boundary reach $\yd$, the hydrogen depletion depth (Equation \ref{eq:yd}). This is the beginning of the event which we have dubbed the ``collision''. The second column of Figure ~\ref{fig:panels} shows the state of the fluid 30 ms later, once the convection zone has fully crossed $y=\yd$. In the collision, fresh fuel in the form of free protons becomes available and triggers CNO burning. A thin burning layer appears at $\yd$ where protons are captured onto $\iso{12}{C}$ previously made from helium burning, and then onto the fresh $\iso{13}{N}$. The protons that do not burn at this boundary are able to travel deeper into the convection zone, but not very far because the temperature is so high. Even though carbon is burning away at the boundary, more is being supplied by helium burning below, and mixing remains efficient enough such that the carbon fraction is uniform throughout the convective region.

With the addition of protons into the mix, CNO burning overtakes helium-burning as the main energy source. One way to track the total energy production in the grid, using horizontally averaged quantities, is with the ``nuclear luminosity'',
\begin{equation}\label{eq:Lnuc}
    L_{\rm nuc}(t)=4\pi R^2\int_z \epsilon_{\rm nuc}(z',t)\rho(z',t) dz'\,,
\end{equation}
where the integral is performed over the vertical direction in the grid. This quantity gives the total energy produced by the burst, at all depths, per unit time\footnote{The actual radiative luminosity coming out of the atmosphere during the burst is much smaller, both because of the thermal time to the surface, and because much of the energy will get used to drive mass loss.}. We plot $L_{\rm nuc}$ in the top panel of Figure~\ref{fig:Lnuc}. We see that after 50 ms, protons initially capture onto $\iso{12}{C}$ to produce $\iso{13}{N}$, which subsequently burns either via proton or $\alpha$-capture. Burning to heavier species such as $\iso{17}{F}$ begins ${\sim}50$ ms later. This sequence of reactions is very similar to what we have previously obtained in 1D simulations.  However, this onset of CNO burning is much weaker than in 1D. This is because the carbon mass fraction at the moment of collision is only $\approx 10^{-3}$, about 100 times smaller than in our \mesa{} simulations. With our initial model, carbon can only come from the tens of milliseconds of helium burning since the start of the simulation, whereas the \mesa{} model already has a significant amount of carbon at ignition (Section~\ref{sec:initial_model}). Due to the limited CNO burning, the convection zone keeps expanding upward in a smooth fashion, albeit at an accelerated rate. This is shown in the bottom panel of Figure~\ref{fig:Lnuc}.

\textit{Growth into the H shell.} Later, as the convection zone, which is becoming richer in heavy elements, keeps growing into the hydrogen-rich shell, it does so fighting against a buoyancy barrier. At this stage, composition gradients start playing an important role in setting the convective boundary. This can be seen in Figure~\ref{fig:T_evolution} with the Ledoux and Schwarzschild boundaries becoming separate. In order to grow, the convection zone has to ``wait'' until it is hot enough that it is buoyant in the light material above. Therefore, a strong temperature gradient develops at the convective boundary. This was predicted by \citet{Weinberg.Bildsten.ea2006}. The material in this gradient region is thermally unstable, but overall convectively stable due to the composition gradient. These are the right conditions for semiconvection to develop \citep{Kippenhahn.Weigert.ea2012}, shown as the blue regions in the bottom panel of Figure~\ref{fig:Lnuc}. To model this type of mixing accurately would require following thermal diffusion and also diffusive mixing of species, which \maestroex{} does not include. Note however that we do not expect either thermal diffusion nor semiconvection to prevent the existence of this sharp temperature gradient. The diffusion coefficient for radiation is $4acT^3/(3\kappa\rho^2c_p)\sim (10^3$ cm$^2$ s$^{-1}$) $T_8^3/\rho_5^2$, and thus thermal time over the ${\sim}10$~cm gradient is $\sim$0.1~s. The diffusion coefficient for semiconvection is similar or even smaller by a factor of up to 1000 \citep{Langer1991}.

During this period, the flow is dominated by large vortices (see top right panel of Figure~\ref{fig:panels}), which is a consequence of the inverse energy cascade of 2D turbulence. Such large vortices are not present in three-dimensional (3D) simulations of X-ray bursts \citep{Zingale.Malone.ea2015}. Interestingly, these vortices appear to carry within them a larger amount of hydrogen than the surrounding mean flow, and are therefore burning at a faster rate than the horizontal mean. Whether the presence of large vortices meaningfully affects the transport of hydrogen and therefore the depth at which burning is taking place is a question which can only be answered by 3D simulations using the same setup. We leave this to future work.

Toward the end of the simulation, carbon and nitrogen have been fully converted to heavier elements, and are no longer available to capture protons, which are still inflowing from the top. The main proton-capture reaction happening at this point, $\iso{16}{O}(p,\gamma)\iso{17}{F}$, is competing with its photodisintegration inverse. As a result, an equilibrium mass fraction of ${\sim}10^{-3}$ of  hydrogen is left behind, which is seen in the third row of Figure~\ref{fig:panels}. We note however that this is likely just an effect of the limited size of our network---these protons would in reality find other avenues to burn.

Throughout the simulation, a secondary convection zone has been developing in the iron substrate underneath the burning layer, as can be clearly seen in the bottom panel of Figure~\ref{fig:Lnuc}. This is the convective \textit{undershoot} problem, previously noted by \citet{Malone.Nonaka.ea2011}, where down currents in the convective region are able to penetrate the stable layer despite opposing buoyancy (which was even artificially enhanced by setting the temperature of the iron substrate to a small value, Section \ref{sec:initial_model}). Chemical mixing between the layers occurs, in which helium and carbon are brought down and a large amount of iron gets dredged up into the burning layer (up to a mass fraction of ${\sim}50$\% toward the end of the simulation), which inhibits the burning. We discuss the validity and potential observational consequences of this undershoot in Section~\ref{sec:discussion}.

Finally, we remark that gravity waves in the stable region are being excited by convection. They can be seen in the vorticity panels of Figure~\ref{fig:panels} (and more clearly in the animated figure). It would be interesting to characterize these waves and their spectrum, but it is beyond the scope of this present study.

\newpage
\subsection{Convective boundary mixing}\label{subsec:boundary}
In \citet{Guichandut.Cumming2023}, the interaction between the convection zone and stable layer at the moment of collision caused an explosive transient with convective velocities at the boundary suddenly increasing, and the convection zone itself extending by many scale heights in the space of a few microseconds. It was unclear if these effects were numerical artifacts, caused by the approximate treatment of convection in the regime of rapid heating. This convective-reactive interaction is an example of a convective boundary mixing process, which are an active area of research (see \citeauthor{Anders.Pedersen2023} \citeyear{Anders.Pedersen2023} for a recent review). Now, we have the necessary fluid simulations to describe what really happens at the top of the convection zone.

Starting with a simple question: How does the convection zone (CZ) grow into the radiative zone (RZ)? The simplest model goes as follows: at any point in time, convection is restricted to the adiabatic region, whereas the RZ above is unchanging. Therefore, the way to move the convective boundary upward is by raising the adiabat of the CZ, i.e. increasing its entropy. This is of course accomplished by nuclear reactions which release heat into the mixture. As such, the convective boundary is always located at the intersection between the initial entropy profile of the radiative zone, $s_{\rm RZ,0}$\footnote{The entropy of the radiative zone does increase over time, but much slower than the evolution of the burst, see \citet{Hanawa.Sugimoto1982}.}, and the \textit{current} entropy of the convection zone, $s_{\rm CZ}=s_{\rm CZ,0}+\Delta s_{\rm nuc}$. This is well illustrated  in Figure 4 of \citet{Hanawa.Sugimoto1982}.

It turns out that this simple picture is not at all what is happening in our simulations. Surprisingly, we find that the specific entropy of the CZ is in fact slowly \textit{decreasing} over time. On its own, this behavior is easily explained by the dredge-up of iron from below, which increases the mean molecular weight faster than the temperature increases from nuclear burning. Nevertheless, it hints at another mechanism for the growth of the CZ. The answer starts in the inset of Figure~\ref{fig:T_evolution}. We see that above the convective boundary, the fluid is becoming cooler than the initial RZ. Heat is being transported away from this region by a mechanism which we describe below. This reduces the entropy of the RZ, allowing it to match to the CZ at smaller and smaller column depths---this is how the CZ grows.

\begin{figure}
    \centering
    \includegraphics[width=0.5\textwidth]{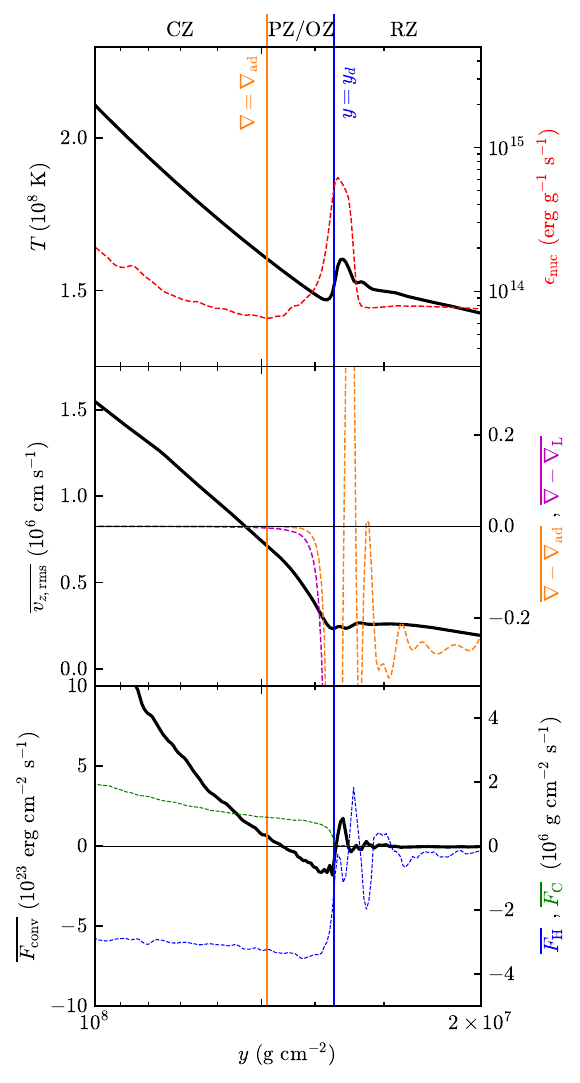}
    \caption{Bulk properties of the flow around the interface between the convective and stable zone prior to the collision, i.e.~a moment where this boundary is below the hydrogen layer (at a column depth $y>\yd$, the vertical blue line). In each panel, the solid black line and the color dashed lines are scaled on the left-hand and right-hand sides respectively. Every quantity is horizontally averaged over the domain and each height. Overbarred quantities are time-averaged over 2 ms. \textit{Top}: temperature profile and burning rate. \textit{Middle}: rms velocity fluctuation and adiabatic excess. The crossing of the dashed orange with zero sets the location of the vertical orange line (Schwarzschild boundary). \textit{Bottom}: convective heat flux and composition fluxes of $\iso{1}{H}$ and $\iso{12}{C}$.}
    \label{fig:boundary_mixing}
\end{figure}

In Figure~\ref{fig:boundary_mixing}, we look at different properties of the flow near the convective boundary, at $t=45$ ms. In the top panel, we notice a temperature bump above the convective boundary. This bump is also clearly seen in Figure~\ref{fig:T_evolution}. The origin of this temperature excess is a spike in burning at the same location, as seen by the red dashed line. This is the same spike which appears in the second row, middle panel of Figure~\ref{fig:panels}, and is a result of the proton captures onto $\iso{12}{C}$ and $\iso{13}{N}$. Indeed, in the middle of Figure~\ref{fig:boundary_mixing}, we see that the rms of $v_z$, the vertical component of the convective velocity, does not go to zero at the convective boundary; despite strong deceleration, some amount of fluid is crossing the boundary and reaching all the way to the hydrogen layer at $y=\yd$. This is how carbon, generated from helium burning at the bottom of the CZ, is able to travel to the hydrogen layer. This is confirmed by the third panel in which we show the horizontal and time average of the carbon flux $F_C\equiv\rho v_z X(\iso{12}{C})$, which stays positive all the way to $\yd$. In the other direction, the returning downflows (with $v_z<0$) preferentially carry with them hydrogen from the top, which results in a negative hydrogen flux $F_H$. As mentioned in the previous section, hydrogen does not burn only at the boundary; much of it is able to travel downward many scale heights. We emphasize that this is all \textit{before} the CZ has actually reached the hydrogen layer---due to these overshooting motions, the collision is initiated earlier than prescribed by the Schwarzschild criterion.

Let us now return to the issue of cooling above the convective boundary. In the middle panel of Figure~\ref{fig:boundary_mixing}, we see that $\nabla$ remains just under $\nabla_{\rm ad}$ over a short distance (about $20\%$ of the local scale height), before dropping, as the temperature gradient goes to the initial radiative gradient $\nabla_{\rm rad}$. This structure is well understood in theories of convective boundary mixing. The marginally subadiabatic region above the Schwarzschild boundary is known as the penetrative zone (PZ), while the radiative region above with nonzero vertical velocity is the overshoot zone (OZ) \citep{Anders.Pedersen2023}. In both regions, upflows are expanding adiabatically in a subadiabatic background, thereby cooling their surroundings. As a result, the convective heat flux, $F_{\rm conv}\equiv \rho c_p v_z \Delta T$, where $\Delta T$ is the temperature difference relative to the horizontal mean, is negative, as seen in the third panel of Figure~\ref{fig:boundary_mixing}\,\footnote{As a quick check, the heat capacity at $y_d$ is ${\approx}10^7$ erg g$^{-1}$. The timescale to reproduce the observed cooling of ${\approx}10^7$ K is $c_p\Delta T/(F/y)\sim 10$~ms, which is consistent with our results.}. Negative heat fluxes in the PZ have been observed in various simulations of convection \citep[e.g.][]{Hurlburt.Toomre.ea1986,Singh.Roxburgh.ea1994,Browning.Brun.ea2004}, but this is to our knowledge the first time the effect is observed in concert with a growing convection zone. This cooling reduces the entropy in the PZ and allows the underlying CZ to expand outward.

\subsection{Burst with initial carbon}\label{subsec:carbon}

\begin{figure*}[h]
    \centering
    \includegraphics[width=\textwidth]{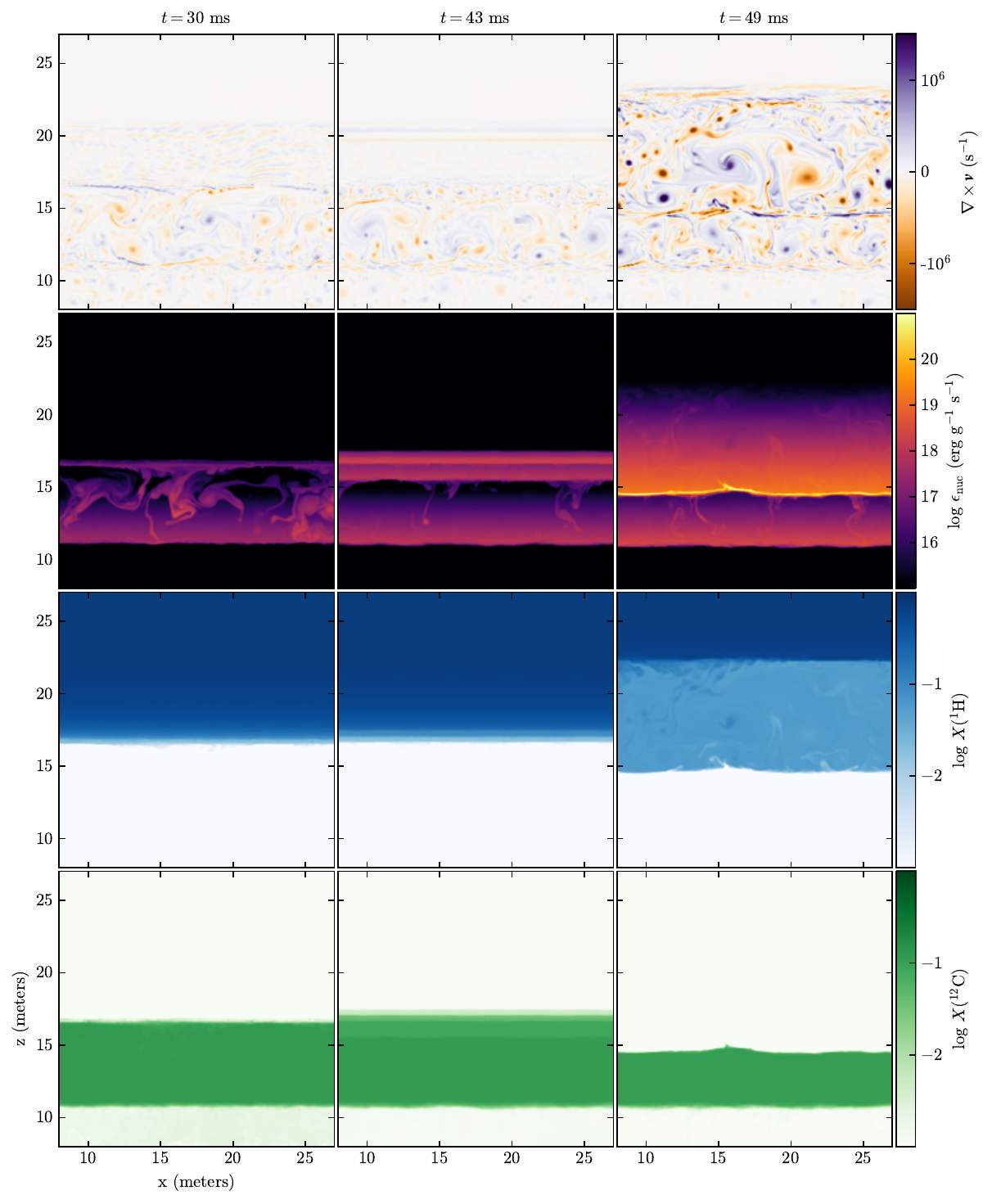}
    \caption{Same as Figure~\ref{fig:panels}, but for the 6 cm resolution including initial carbon (model B). Note that the scales for nuclear burning and carbon mass fractions are different than in Figure~\ref{fig:panels}. We are also showing a larger portion of the box (8--27~m, instead of 8--22~m), to better show the growth of the convection zone. An animated version of this figure is available in the online journal.}
    \label{fig:panels_carbon}
\end{figure*}

Now, we simulate a burst closer to what we had in \mesa{}, which produced a violent collision. We start with the initial model with carbon presented in Section \ref{sec:initial_model}, and run it on a 6 cm per zone grid. The simulation is not fully converged at this resolution according to our previous findings (see Figure~\ref{fig:peaks}). However, our aim in this section is only to discuss important differences in the evolution of the burst. To highlight these differences, we select three new special times and show snapshots of the fluid in Figure~\ref{fig:panels_carbon}.

The early evolution of the burst is similar to before, with triple-$\alpha$ dominating the energy release. Starting at about 30 ms (left column of Figure~\ref{fig:panels_carbon}), the effects of the added carbon are already noticeable. As we discussed in the previous section, even before the collision, overshooting motions above the convective boundary cause mixing between the convective and stable layers. The protons coming down now capture much faster, and CNO burning already becomes the dominant source of nuclear energy (see the first column, second row panel of Figure~\ref{fig:panels_carbon}). This reinforces the idea that the way in which the fuel burns during the burst is very sensitive to the details of mixing at the convective boundary.

At first, burning is slow enough that heat is efficiently distributed throughout the convection zone. As burning becomes faster, the temperature profile begins to increase locally, splitting the convection zone into multiple layers. These layers are well-mixed but have different compositions (see the hydrogen and carbon fractions in the second column of Figure~\ref{fig:panels_carbon}). Interestingly, we had also found this splitting of the convection zone in our \mesa{} simulations, but it was much more severe. Instead of 3-4 individual layers as we have here, \mesa{} was producing on the order of 10--100 layers, depending on resolution. We had dismissed this splitting as an artifact of the approximate treatment of convection with MLT, but these new simulations indicate that it is real effect. We return to this point in Section~\ref{sec:discussion}.

The layers are short-lived. Only a few milliseconds later, in what looks like a secondary and much more powerful runaway, the top three layers merge and drastically expand both up and down (third column of Figure~\ref{fig:panels_carbon}). The downward propagation rapidly consumes the carbon in the bottom convective layer. The burning rate at the interface between the two layers is $\approx10^{21}$ erg g s$^{-1}$, the highest of any of our simulations. The top convective boundary moves from $y\approx y_d$ to ${\approx}\,0.1y_d$ in ${\approx}\,1$~ms. In \mesa{}, this same expansion occurred in a tenth of the time \citep[see Figure 3 of][]{Guichandut.Cumming2023}, so in fact the hydrodynamical collision is not as violent as in our 1D simulations, even with a similar $\iso{12}{C}$ fraction. We speculate that this is because we did not include an overshoot prescription in 1D. As we have seen, overshooting motions cause the proton ingestion to begin early, before the Schwarzschild boundary reaches $y=y_d$, and not all at once. The other reason is that we did not consider acceleration-limited convection in 1D, which would have prevented the fluid from spontaneously reaching large velocities, potentially slowing down the expansion of the convection zone.

In the final stage of rapid expansion, 50 ms after the start of the simulation, the Mach number becomes large, averaging at $\sim8-10$\% and peaking at almost 25\% in the upper convection zone. This is pushing the limits of the low-Mach method. We conclude that a fully compressible calculation is needed to study this type of burst, especially when the collision is explosive.

\section{Summary and Discussion}\label{sec:discussion}

We have run low-Mach number simulations of X-ray bursts in the regime where helium is igniting and the convection zone eventually reaches new fuel in the form of protons, which accelerates the thermonuclear runaway. Our main findings are as follows:

\begin{enumerate}
    \item The resolution required to reach convergence is 3 cm per zone or smaller (Figure~\ref{fig:peaks}). At coarser resolution, the evolution of the burst is delayed due to improper resolving of the temperature peak.
    \item Nuclear burning throughout the burst evolves in a similar fashion to 1D simulations. The burst is initially driven by helium burning. Once protons enter the mix, there is a buildup of $\iso{13}{N}$, which then quickly burns to $\iso{14}{O}$, increasing the nuclear energy output (Figure~\ref{fig:Lnuc}).
    \item The collision is not a single precise moment, but rather a gradual encroaching of the convection zone into the hydrogen shell, which is mediated by convective boundary mixing processes (Figure~\ref{fig:boundary_mixing}). Early on, convective motions penetrate through the convective boundary, exchanging entropy and composition with the stable layer above. A negative heat flux reduces the entropy above the boundary, which allows the convection zone to grow.
    \item The evolution of the burst is highly sensitive to the initial amount of carbon in the layer. When it is high, CNO burning dominates the energy release as soon as protons get entrained into the convection zone. The rapid burning leads to a splitting of the convection zone into separate well-mixed layers, which soon merge again in a violent CNO-driven runaway.
\end{enumerate}

In simulating bursts with two different initial fuel compositions, with and without carbon, we have probed two different points in the accretion/ignition parameter space. If the burst were to ignite soon after hydrogen depletion, it would be in a pure He background, as there would be no time to make carbon. For the set of accretion parameters in our \mesa{} simulation, hydrogen depletion was reached after 10 hours, and ignition occurred 5.5 days later. What fraction of this parameter space leads to violent collisions, and how might this tie to observations? Since hydrodynamical simulations are computationally expensive, this is most likely a question for 1D simulations. We leave this to future work.

Our simulations are not able to follow the full evolution of convection during the burst. This is mainly a dynamic range issue; there is a $\sim$4 order of magnitude difference between the density at the bottom of the fuel layer ($y\sim 10^8$ g cm$^{-2}$, $\rho\sim 10^6$ g cm$^{-3}$) and that of the theoretical maximal extent of the convection zone \citep[$y\sim 10^4$ g cm$^{-2}$, $\rho\sim 10^2$ g cm$^{-3}$, e.g.][]{Weinberg.Bildsten.ea2006}. It is difficult to track this range in the low-Mach approximation because of the need for a velocity-damping region (the sponge) at the top of the model. The smallest value of $\rho_{\rm cutoff}$ which would allow us to evolve the burst for a long time was $10^3$ g cm$^{-3}$. Following the growth of the convection zone into regions of lower density therefore likely requires a fully compressible fluid simulation. These will allow us to constrain the duration of the observed ``pause'' as in \src{} \citep{Bult.Jaisawal.ea2019,Guichandut.Cumming2023}.

Our simulations did not include thermal diffusion. Although this is possible with \maestroex{}, it is not expected to be a big effect in the early stages of the burst, where convection is taking place at high densities and the conductivity is small. However, it is thermal diffusion which ultimately stops the advancement of convection at lower depths, when it becomes more efficient as a means to transport energy than convection. Future simulations will therefore need to include it to describe the full evolution of convection.

In addition, although previous comparisons between 2D and 3D simulations of bursts have demonstrated good overall agreement in the burst evolution \citep{Zingale.Malone.ea2015}, it is well known that the turbulent flow behaves very differently. In this work, we noticed the presence of large vortices which carry a different composition than the surroundings. These may completely disappear in 3D, which might alter the mixing and burning at different heights. We remark that a fully compressible 3D simulation of this burst is within reach with current computational resources, and we have already started some experiments with the \texttt{CASTRO} code \citep{Almgren.Beckner.ea2010} which show a similar early burst evolution as our \maestroex{} simulations. Furthermore, rotation could be important in dictating the flow patterns, as the spin period of accreting neutron stars can be on the order of milliseconds, comparable to the convective turnover time. These effects have recently been explored in the Boussinesq approximation by \citet{Garcia.Chambers.ea2018}, who modeled burning-driven convection over the whole surface of the star.

Another potential impact of dimensionality is in the convective undershoot, which has an important impact on our simulations. As mentioned in Section~\ref{subsec:evolution}, the dredge-up of a large amount of iron into the convective layer reduces the nuclear burning rate. Also, a large fraction of the nuclear energy (about 65\% in the main 3 cm simulation) is used up to raise the temperature of the bottom layer (Figure~\ref{fig:T_evolution}). An interesting comparison can be drawn to the work of \citet{Kercek.Hillebrandt.ea1998}, who simulated 2D thermonuclear runaways on a white dwarf. Similar to us, they found large vortices, heating of the undershoot layer, and a significant dredge-up of the underlying material (see their figures 1, 3b, and 4). In a subsequent 3D study \citep{Kercek.Hillebrandt.ea1999}, these effects were much less severe. In our own preliminary 3D calculations, we also find that undershoot and dredge-up are diminished. This is an observationally relevant question, as dredge-up of heavy elements synthesized in previous bursts could have implications for the detection of spectral features in some photospheric radius expansion bursts \citep{Zand.Weinberg2010,Kajava.Nattila.ea2017,Strohmayer.Altamirano.ea2019}

Lastly, a complete treatment of Type I X-ray bursts requires not only the vertical but also lateral heat transport around the surface of the neutron star. This is the well-studied flame-propagation problem, which has been applied to pure He \citep[e.g.][]{Cavecchi.Watts.ea2013,Eiden.Zingale.ea2020,Chen.Zingale.ea2023} and most recently to mixed H/He atmospheres \citep{Johnson.Zingale2024}, but not yet to the H/He layered case of this paper. How does our concept of ``collision'' fit into this alternate picture? Is the flame fast enough to propagate before convection has had time to connect with the hydrogen shell, or do these processes interact in some other way? We hope our work motivates future explorations into this question.\\

\maestroex{} is open-source and under active development. The code as well as the inputs required to run the simulations presented in this paper can be found at \url{https://github.com/amrex-Astro/Maestroex}.

We thank J.R. Fuentes for useful discussions on convective boundaries, E.T. Johnson and A. Smith Clark for technical assistance. We also thank the referee for useful comments that led us to appreciate the important role of the initial amount of carbon. This project was borne out of a visit at Stony Brook University, thanks to funding by the Centre de Recherche en Astrophysique du Québec (CRAQ). S.G. is also supported by an NSERC scholarship. The work at Stony Brook was supported by DOE/Office of Nuclear Physics grant DE-FG02-87ER40317. This work was supported by NSERC Discovery Grant RGPIN-2023-03620. This research used resources of the National Energy Research Scientific Computing Center (NERSC), a Department of Energy Office of Science User Facility using NERSC award NP-ERCAP0027167.

\software{\maestroex{} \citep{Nonaka.Almgren.ea2010}, \texttt{AMReX}
\citep{Zhang.Almgren.ea2019}, \texttt{pynucastro} \citep{Smith-Clark.Johnson.ea2023}, \texttt{NumPy} \citep{Harris.Millman.ea2020}, \texttt{SciPy} \citep{Virtanen.Gommers.ea2020}, \texttt{Matplotlib} \citep{Hunter2007}, and \texttt{yt} \citep{Turk.Smith.ea2011}.}
\facilities{NERSC}

\bibliographystyle{aasjournal}
\bibliography{references} 

\appendix 
\section{Building the initial model}\label{sec:app_initial_model}

Our model is an extension of the one used in \citet{Malone.Zingale.ea2014}, in which the main feature is an isentropic zone in the fuel layer which becomes convectively unstable once the simulation begins. For this, we provide the temperature of the ``neutron star'' $T_{\rm star}$, i.e. the temperature in the iron substrate, the height $H_{\rm star}$ of this substrate, the temperature and column depth of the base of the fuel layer, $T_{\rm base}$ and $y_{\rm base}$. At $r=H_{\rm star}$, the composition transitions from pure iron to the fuel composition with a hyperbolic tangent of width $\delta$ (see appendix of \citet{Malone.Zingale.ea2014}). We set $\delta=5$ cm so that this transition is thin, but well resolved at the largest grid resolutions.

In \citet{Malone.Zingale.ea2014}, the fuel layer was a mixture of H and He at fixed mass fractions. Here, the composition depends on the radial location. In particular, the mass fraction of hydrogen $X$ depends analytically on the column depth $y$ \citep{Cumming.Bildsten2000},
\begin{equation}\label{eq:X(y)}
    X(y)=\rm{max}(0,\,X_0[1-y/\yd])\,
\end{equation}
where $X_0$, the initial hydrogen fraction, and $\yd$, the hydrogen depletion depth, are two more input parameters. At any point during the building of the model, $y$ can be obtained from the pressure, since $p=gy$ in HSE. For the CNO species, we set constant mass fractions of $\iso{14}{O}$ and $\iso{15}{O}$, with a sum of $Z_{\rm O}$, and a ratio of 70.621/122.266, which is the ratio of beta-decay half-lives of these isotopes \citep{Kondev.Wang.ea2021}.  Finally, the He mass fraction is simply $Y=1-X-Z_{\rm O}$.

For the alternate model with an initial amount of carbon (model B), we fit a power law of the form $Z_C(y)=(y/y_C)^{\nabla_C}$ to the $\iso{12}{C}$ mass fraction in the \mesa{} model. We tweak $y_C$ such that the total mass of carbon is within 5\% of the \mesa{} model. Because there is slightly less fuel than in the \mesa{} model, this condition requires a slightly larger initial carbon fraction (compare dash-dotted and dotted green lines in Figure~\ref{fig:initial_models}). This is done so that, once the convection zone is mixed, the uniform carbon fraction is similar between the two models (about 10\%). The oxygen mass fractions are kept the same as before, and $Y=1-X-Z_{\rm O}-Z_C$.

\begin{figure*}[h]
    \centering
    \includegraphics{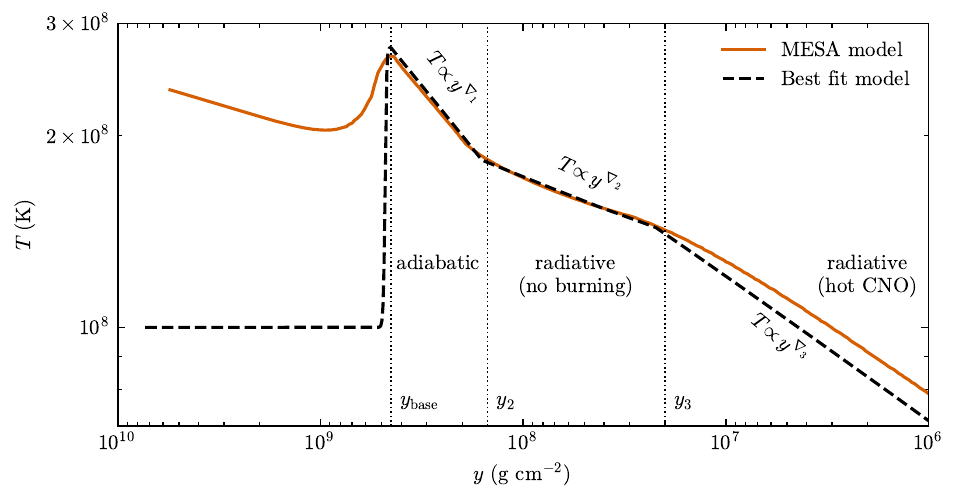}
    \caption{Illustration of the temperature profile considered for the toy model. In each zone demarcated by the vertical dotted lines (except for the isothermal substrate at large depths $y>y_{\rm base}$), the best-fitting temperature gradient $\nabla=d\ln T/d\ln y$ is obtained from the \mesa{} model. This temperature profile is then fed into a hydrostatic solver, which determines the density at every point, while the composition is an analytical function of the pressure with Equation~\ref{eq:X(y)}. See text for further details.}
    \label{fig:Tprof_illustration}
\end{figure*}

The initial temperature profile is also more complicated than in \citet{Malone.Zingale.ea2014}. Above the isentropic zone where the temperature gradient,$\nabla\equiv d\ln T/d\ln y$, is the adiabatic gradient $\nabla_{\rm ad}$, the atmosphere is in radiative equilibrium, $dT^4/dr\propto -F$, where $F$ is the radiative flux, which depends on the depth and nuclear burning from steady-state energy balance, $dF/dy=-\epsilon_{\rm nuc}$. In practice, instead of solving for an exact temperature profile consistent with the burning, we divide the fuel layer into three zones and obtain the best-fitting value of $\nabla$ in each zone from the \mesa{} model. This is illustrated in Figure~\ref{fig:Tprof_illustration}. On top of the three $\nabla$'s, the column depth boundaries between the first and second zones, $y_2$, and second and third zones, $y_3$, also must be specified as input parameters.

In 1D calculations, ignition is manifested by a rapid reduction in the time step which is limited by the nuclear energy generation rate. However, in a fluid simulation, the time step is restricted from the start by the fluid velocity (Courant condition). Therefore, a simulation starting at this point will take a long time (many time steps) to evolve and proceed into the thermonuclear runaway. In order to accelerate it, we ``kick'' the initial model, i.e. we artificially increase the temperature at the base of the fuel layer ($T_{\rm base}$ parameter), while keeping the temperature gradient in that region $\nabla_1$ constant. As a result, the boundary of the adiabatic region moves according to
\begin{equation}
    y_2'=y_2\left(\frac{T_{\rm base}}{T_{\rm base}'}\right)^{(\nabla_1-\nabla_2)^{-1}}\,,
\end{equation}
where the prime ($'$) values are those of the kicked model.

Table~\ref{tab:input_params} below lists all of the parameters used to build the initial fluid model.

\begin{table*}[h]
    \begin{center}
    \caption{Input parameters for initial model}
    \label{tab:input_params}
    \hspace*{-1cm}\begin{tabular}{cccc}
        \hline\hline
        Parameter & Definition & Value & Units \\
        \hline
         $T_{\rm star}$ & Temperature at the bottom of the atmosphere & $10^8$ & K\\
         $H_{\rm star}$ & Height of the iron substrate & 1070 & cm\\
         $\delta$ & Width of the hyperbolic tangent transition between iron and fuel & 5 & cm\\
         $T_{\rm base}$ & Temperature at the bottom of the fuel layer & $3.5\times 10^8\,^{\rm (a)}$ & K\\
         $\nabla_1$ & Temperature gradient in the He isentropic zone & $0.37\,^{\rm (b)}$ & \\
         $\nabla_2$ & Temperature gradient in the inert He zone & 0.12 & \\
         $\nabla_3$ & Temperature gradient in the H-rich zone & 0.22 & \\
         $y_{\rm base}$ & Column depth at the base of the fuel layer & $4.5\times 10^8$ & g cm$^{-2}$\\
         $y_2$ & Column depth boundary between the $\nabla_1$ and $\nabla_2$ zones & $5.6\times 10^7\,^{\rm (a)}$ & g cm$^{-2}$\\
         $y_3$ & Column depth boundary between the $\nabla_2$ and $\nabla_3$ zones & $2\times 10^7$ & g cm$^{-2}$\\
         $X_0$ & Hydrogen fraction in the accreted material & 0.7 & \\
         $Z_{\rm O}$ & Total oxygen ($\iso{14}{O}+\iso{15}{O}$)  mass fraction in the accreted material & 0.02 & \\
         $\yd$ & Hydrogen depletion depth from stable hot CNO burning & $3.7\times 10^7$ & g cm$^{-2}$\\
         $y_C$ & Column depth parameter for the carbon fraction$\,^{\rm (c)}$ & $5.9\times 10^8$ & g cm$^{-2}$\\
         $\nabla_C$ & Power law index for the carbon fraction$\,^{\rm (c)}$ & $3.16$ & \\
         $T_{\rm cutoff}$ & Minimum temperature in the model & $10^6$ & K\\
         $\rho_{\rm cutoff}$ & Minimum density in the model & $10^3$ & g cm$^{-3}$\\
         \hline
    \end{tabular}
    \end{center}
    \hspace*{1.5cm}$^{\rm (a)}$ These are the values used for the kicked model, i.e. the one shown in Fig.~\ref{fig:initial_models}, not Fig.~\ref{fig:Tprof_illustration}.\\
    \hspace*{1.5cm}$^{\rm (b)}$ This is approximately the correct value of $\nabla_{\rm ad}$ in that region.\\
    \hspace*{1.5cm}$^{\rm (c)}$ These parameters are ignored in the case of model A (which does not include carbon).\\
\end{table*}

\end{document}